\documentclass[graybox]{svmult}
\usepackage{epsfig}
\usepackage{graphicx}
\usepackage{amsfonts}

\usepackage{epsfig}
\usepackage{mathptmx}   
\usepackage{helvet}     
\usepackage{courier}    
\usepackage{type1cm}    
\usepackage{makeidx}    
\usepackage{graphicx}   
\usepackage{multicol}   
\usepackage[bottom]{footmisc}   

\begin{document}

\title{Defect production due to quenching through a multicritical point and 
along a gapless line}
\author{Uma Divakaran$^{(1)}$, Victor Mukherjee$^{(1)}$, Amit Dutta$^{(1)}$ 
and Diptiman Sen$^{(2)}$}

\institute{$^{(1)}$ Department of Physics, Indian Institute of Technology 
Kanpur, Kanpur 208 016, India \\
$^{(2)}$ Center for High Energy Physics, Indian Institute of Science,
Bangalore 560 012, India}

\titlerunning{Defect production due to quenching through $\cdots$}
\maketitle

\section{Introduction}

The exciting physics of quantum phase transitions has been explored 
extensively in the last few years \cite{dmdssachdev99,dmdsdutta96}. 
The non-equilibrium dynamics of a quantum system when quenched very fast 
\cite{dmdssengupta04} or slowly across a quantum critical point
\cite{dmdszurek05,dmdspolkovnikov05} has attracted the attention of several 
groups recently. The possibility of experimental realizations of quantum
dynamics in spin-1 Bose condensates \cite{dmdssadler06} and atoms trapped 
in optical lattices \cite{dmdskitaev_exp,dmdsbloch07}
has led to an upsurge in studies of related theoretical models 
\cite{dmdssengupta04,dmdszurek05,dmdspolkovnikov05,dmdskadowaki,dmdscalabrese,
dmdsdziarmaga05,dmdsdamski06,dmdslevitov06,dmdsdziarmaga06,dmdscucchietti07,
dmdsmukherjee07,dmdsdivakaran07,dmdsosterloh07,dmdspolkovnikov07,
dmdssengupta08,dmdssen08,dmdssantoro08,dmdspatane08,dmdsmukherjee08,
dmdsdivakaran08,dmdscaneva08,dmdsviola08,dmdsdziarmaga08,dmdsbermudez08,
dmdsdivakaran09,dmdsmukherjee09,dmdsdivakaran091}.
 
In this review, we concentrate on the dynamics of quantum spin chains swept 
across a quantum critical or multicritical point or along a gapless line by 
a slow (adiabatic) variation of a parameter appearing in the
Hamiltonian of the system. 
Our aim is to find the scaling form of the density of defects (which, 
in our case, is the density of wrongly oriented spins) in the final
state which is reached after the system is prepared in an initial ground 
state and then slowly quenched through a quantum critical point.
The dynamics in the vicinity of a quantum critical point is necessarily 
non-adiabatic due to the divergence of the relaxation time of the underlying 
quantum system which forces the system to be infinitely sluggish; thus the 
system fails to respond to a change in a parameter of the
Hamiltonian no matter how slow that rate of change may be!

We first recall the Kibble-Zurek argument \cite{dmdskibble76,dmdszurek96} 
which predicts a scaling form for the defect density following a slow quench 
through a quantum critical point. We assume that a parameter $g$ of the 
Hamiltonian is varied in a linear fashion such that $g-g_c \sim t/\tau$, 
where $g=g_c$ denotes the value of $g$ at the quantum critical point and 
$\tau$ is the quenching time. Our interest is in the
adiabatic limit, $\tau \to \infty$. The energy gap of the quantum 
Hamiltonian vanishes at the critical point as $(g-g_c)^{\nu z}$ whereas 
the relaxation time $\xi_{\tau}$, which is inverse of the gap, diverges at 
the critical point. It is clear that non-adiabaticity becomes important at 
a time $\hat t$ when the characteristic time scale of the quantum system 
(i.e., the relaxation time) is of the order of the inverse of the rate 
of change of the Hamiltonian; this yields
\begin{equation} \hat t \sim \xi_{\tau}(\hat t)\rightarrow \hat t \sim
(g-g_c)^{-\nu z} \sim \left(\frac{\hat t}{\tau}\right)^{-\nu z}. 
\nonumber \\ \end{equation}
This leads to a characteristic length scale $\xi$ given by
\begin{equation} \xi \sim \tau^{-\nu/(\nu z+1)}, \end{equation}
where $\xi$ is the correlation length or healing length. The healing length 
typically denotes the length over which a single defect is present. The 
density of defects in a $d$-dimensional system scales as $1/\xi^d$ which 
leads to the Kibble-Zurek scaling form for the density of defects $n$ given by
\begin{eqnarray} n\sim \tau^{-d \nu/(\nu z+1)}. \label{dmdsdensity1} 
\end{eqnarray}
The interesting aspect of the Kibble-Zurek prediction is that the scaling form
of the defect density in the final state of a driven quantum system varies in 
a power-law fashion with the rate of quenching $1/\tau$, and the exponent of 
the power-law depends on the spatial dimension $d$ and the static
quantum critical exponents $\nu$ and $z$. 

The Kibble-Zurek scaling has been verified in various exactly
solvable spin models and systems of interacting bosons 
\cite{dmdszurek05,dmdslevitov06,dmdsmukherjee07,dmdsdivakaran07,
dmdsdziarmaga05,dmdspolkovnikov05}; it has been generalized to quenching 
through a multicritical point \cite{dmdsdivakaran09}, across a gapless 
phase \cite{dmdssengupta08,dmdssantoro08}, and along a gapless line 
\cite{dmdsdivakaran08,dmdsviola08}, and to systems with quenched disorder 
\cite{dmdsdziarmaga06}, white noise\cite{dmdsosterloh07}, 
infinite-range interactions \cite{dmdscaneva08}, and edge states 
\cite{dmdsbermudez08}. Studies have also been made to estimate the defect 
density for quenching with a non-linear form \cite{dmdssen08}, an oscillatory 
variation of an applied magnetic field \cite{dmdsmukherjee09} or
under a reversal of the magnetic field \cite{dmdsdivakaran091}. In the 
article in this book by Mondal, Sengupta and Sen \cite{dmdsmondal091}, the 
quenching dynamics through a gapless phase and quenching with a power-law 
form of the change of a parameter as well as the possibility of experimental 
realizations have been discussed in detail. It should be mentioned that
in addition to the density of defects in the final state, the degree of 
non-adiabaticity can also be quantified by looking at various quantities like
residual energy\cite{dmdskadowaki,dmdssei} and fidelity \cite{dmdszurek05}.

\section{A spin model: transverse and anisotropic quenching}

\begin{figure}[htb]
\begin{center} \includegraphics[height=2.7in,width=3.1in]{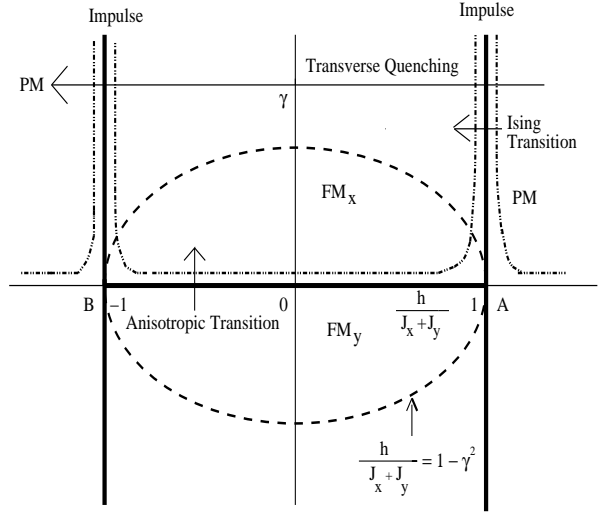}
\caption{The phase diagram of the anisotropic $XY$ model in a transverse 
field in the $~h/(J_x + J_y) ~- ~\gamma ~$ plane, where $\gamma \equiv 
(J_x -J_y)/(J_x + J_y)$. The vertical bold lines given by $~h/(J_x + J_y) 
= \pm 1~$ denote the Ising transitions. The system is also gapless on the 
horizontal bold line $\gamma
= 0$ for $~|h| < J_x + J_y$. ${\rm FM_x ~(FM_y)}$ is a long-range ordered 
phase with ferromagnetic ordering in the $x ~(y)$ direction. The thick dashed 
line marks the boundary between the commensurate and incommensurate 
ferromagnetic phases. The thin dotted lines indicate the adiabatic and impulse
regions when the field $h$ is quenched from $-\infty$ to $\infty$. The two 
points with coordinates $\gamma = 0$ and $h/(J_x+J_y) =\pm 1$ denoted by $A$ 
and $B$ are the multicritical points.} \label{dmdsphase_xy} \end{center} 
\end{figure}

In this article, we will focus mainly on the one-dimensional anisotropic
$XY$ spin-1/2 chain in a transverse field \cite{dmdsbunder99,dmdslieb61} 
which is represented by the Hamiltonian
\begin{eqnarray} H~=~ -~\sum_n ~[J_x \sigma^x_n\sigma^x_{n+1} ~+~ J_y
\sigma^y_n \sigma^y_{n+1} ~+~ h \sigma^z_n], \label{jxy} \end{eqnarray}
where $\sigma's$ are the usual Pauli matrices. (We will set Planck's
constant $\hbar =1$). The spectrum of this
Hamiltonian can be found exactly by first mapping the $\sigma$ matrices to
Jordan-Wigner fermions $c_n$ as \cite{dmdslieb61}
\begin{eqnarray} c_n &=& \sigma_n^- ~\exp(i\pi\sum_{j=1}^{n-1}
\sigma_j^+ \sigma_j^-) , \nonumber \\
\sigma^z_n &=& 2c_n^{\dagger} c_n - 1, \end{eqnarray}
where $\sigma_n^{\pm} = \sigma_n^x \pm i \sigma_n^y$ are spin raising and 
lowering operators respectively. These Jordan-Wigner fermion operators follow 
the usual anticommutation rules
\begin{eqnarray} \{c_m^{\dagger},c_n\}=\delta_{mn}; ~~~~\{c_m,c_n\}=0 =
\{c^{\dagger}_m,c^{\dagger}_n\}. \end{eqnarray}
Applying a Fourier transformation to the Jordan-Wigner fermions along 
with the periodic boundary conditions, the Hamiltonian in (\ref{jxy})
can be rewritten as 
\begin{eqnarray} H &=& - ~\sum_{k>0} ~\{ ~[(J_x + J_y) \cos k +h] ~
(c_k^{\dagger} c_k + c_{-k}^{\dagger} c_{-k}) \nonumber \\
& & ~~~~~~~~~~~~~+ i (J_x -J_y) \sin k ~(c_k^{\dagger} c_{-k}^{\dagger}
- c_{-k} c_k \}, \label{dmdsh2} \end{eqnarray}
where $k$ lies in the range $[0,\pi]$.
The Hamiltonian is quadratic in fermion operators and hence exactly solvable 
using an appropriate Bogoliubov transformation \cite{dmdslieb61}.
Using the basis vectors $|0 \rangle$ (where no fermions are present) and 
$|k,-k \rangle = c_k^{\dagger} c_{-k}^{\dagger} |0 \rangle $, we can recast 
the Hamiltonian for wave number $k$ in a $2\times 2$ matrix form
\begin{equation} H_k=\left[\begin{array}{ll}
h+(J_x+J_y)\cos k& ~~~~i (J_x-J_y)\sin k \\
-i(J_x-J_y)\sin k&~~~~-h-(J_x+J_y)\cos k\end{array}\right] , \label{dmdsmatrix}
\end{equation}
The spectrum for this is given by
\begin{equation} \epsilon_k ~=~ [~h^2 +J_x^2 + J_y^2 + 2 h (J_x + J_y) \cos k 
+ 2 J_x J_y \cos 2k ~]^{1/2}. \label{dmdsek} \end{equation}
This Hamiltonian has a very rich phase diagram as shown in Fig. 
\ref{dmdsphase_xy}. The vanishing of the energy gap for $h=\pm (J_x+J_y)$ 
for the critical modes at $k=\pi$ and $0$, respectively, signals quantum 
phase transitions from a ferromagnetically ordered phase to a paramagnetic 
phase with critical exponents $\nu=z=1$ \cite{dmdsbunder99}. On the other 
hand, the vanishing energy gap at $J_x=J_y$ and $h<J_x+J_y$ denotes the 
anisotropic transition which marks the boundary between the two ferromagnetic 
phases with the critical exponents $\nu$ and $z$ being identical to the Ising 
transition. The meeting points of these two transition lines at $h= \pm 
(J_x+J_y)$ and $J_x=J_y$ are multicritical points.

Let us initiate our discussions with two types of quenching schemes: 

\noindent (i) quenching the magnetic field $h$ as $t/\tau$ which we call 
transverse quenching \cite{dmdsdziarmaga05,dmdslevitov06}, and 

\noindent (ii) the quenching of the interaction $J_x$ as $t/\tau$ which is 
referred to as anisotropic quenching \cite{dmdsmukherjee07}. 

\noindent
In the process of anisotropic quenching, the system can be made to 
cross the multicritical points $A$ and $B$ shown in the phase diagram. The 
quenching through a multicritical point is a cardinal point of our discussion.
We shall also discuss the quenching scheme where the anisotropy parameter 
$\gamma=J_x-J_y$ is quenched in a linear fashion keeping the system always 
on the gapless line $h=J_x+J_y$. It will be shown that
in either case, we arrive at new scaling behaviors for the defect 
density which cannot be obtained by a simple fine tuning of the 
Kibble-Zurek scaling form $n\sim 1/\tau^{\nu d/\nu z+1}$.

\begin{figure}[htb]
\begin{center} \includegraphics[height=2in]{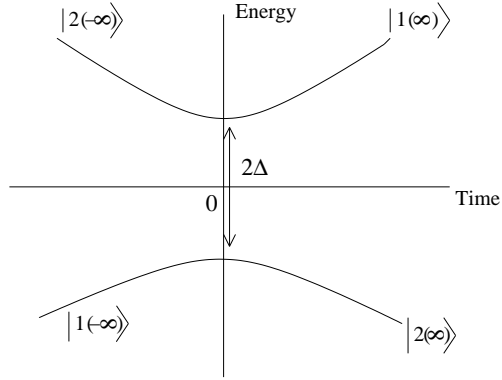}
\caption{The two levels correspond to the energy levels varying with time; 
$2\Delta$ is the minimum gap between these two levels, and $|1 \rangle$ and 
$|2 \rangle$ are the eigenstates in the asymptotic limits $t \to \pm \infty$.
The ground state changes its characteristics from $|1 \rangle$ at $t \to -
\infty$ to $|2 \rangle$ at $t\to +\infty$.} \label{dmdslandau} \end{center}
\end{figure}

The reduction of the general Hamiltonian to a direct product of 
$2\times 2$ matrices as given in (\ref{dmdsmatrix}) facilitates the 
application of the Landau-Zener (LZ) transition formula \cite{dmdslandau} 
to estimate the non-adiabatic transition probability on passing through a 
quantum critical point. The general LZ Hamiltonian given by
\begin{equation} H ~=~ \epsilon_1 |1 \rangle \langle 1| ~+~ \epsilon_2 |2 
\rangle \langle 2| ~+~ \Delta \left( |1 \rangle \langle 2| ~+~ |2 \rangle 
\langle 1| \right) \end{equation}
closely resembles the reduced $2 \times 2$ spin Hamiltonian given in 
(\ref{dmdsmatrix}). In the LZ Hamiltonian, we set $\epsilon_1 -\epsilon_2= 
t/\tau$. The two energy levels $\pm \sqrt{\epsilon^2+ \Delta^2}$, where 
$\epsilon_1 =-\epsilon_2 =\epsilon$, approach
each other with a minimum gap $2 \Delta$ at $t=0$ as shown in Fig. 
\ref{dmdslandau}. The system is prepared in its initial ground state
$|1 \rangle$ at time $t \to -\infty$ and should reach the final ground state 
$|2 \rangle$ at $t \to +\infty$ if the dynamics is adiabatic throughout. 
A general state during the time evolution can be written in the form
$| \psi(t) \rangle = C_1 (t) |1 \rangle + C_2(t) |2 \rangle$ with the initial 
condition $|C_1(t\to -\infty)|^2 =1$. The LZ probability for the non-adiabatic 
transition is given by \cite{dmdslandau,dmdssei}
\begin{eqnarray} p= |C_1(t \to \infty)|^2= e^{-2\pi \Delta^2/|
\frac{\partial}{\partial t} (\epsilon_1-\epsilon_2) |}. \end{eqnarray}
It is to be noted that the above formula is valid for a linear variation 
of the bare levels $\epsilon_1$ and $\epsilon_2$ and also when the 
off-diagonal term does not include any time dependence.

Let us now discuss the transverse quenching of the $XY$ spin chain discussed 
above. The phase diagram in (1.1) shows that when the transverse field is 
quenched from $-\infty$ to $+\infty$, the system is swept across the
two Ising transition lines at $h =\pm (J_x + J_y)$. 
Referring to the Hamiltonian in (\ref{dmdsmatrix}), it is clear that when $h$ 
is quenched as $h\sim t/\tau$ \cite{dmdslevitov06}, the dynamics of 
the spin chain effectively reduces to a LZ problem in the two-dimensional 
reduced Hilbert space spanned by the basis
vectors $|0 \rangle$ and $|k,-k \rangle$, with the initial ground state being 
$|0 \rangle$. Here, the diagonal and off-diagonal terms are denoted by
$\epsilon_k^{\pm} = \pm [h+(J_x+ J_y) \cos k]$ and $\Delta = i 
(J_x -J_y) \sin k$ respectively. 
Denoting a general state vector $\psi_k(t)$ at an instant $t$ as $|\psi_k(t) 
\rangle = C_{1,k} |0 \rangle + C_{2,k} |k,-k \rangle$ with $|C_{1,k}
(t\to -\infty)|^2=1$, the non-adiabatic transition probability for the mode
$k$ is directly obtained using the LZ transition formula 
\begin{eqnarray} p_k= |C_{1,k}(t \to \infty)|^2 = e^{-\pi \tau |J_x - J_y|^2 
\sin^2 k}. \end{eqnarray}
The density of defects (i.e., density of wrongly oriented spins) in the 
final state is obtained by integrating $p_k$ over the Brillouin zone 
\cite{dmdslevitov06},
\begin{equation} n=\frac{1}{2\pi}\int_{-\pi}^{\pi} p_k dk \approx 
\frac 1 {\pi \sqrt {\tau} (J_x -J_y)}. \end{equation}
In the adiabatic limit $\tau \to \infty$, the transition probability is 
non-zero only for modes close to the critical modes $k=0$ and $\pi$; this
allows us to extend the limits of integration from
$-\infty$ to $+\infty$. Noting that the critical exponents $\nu=z=1$ for 
the Ising transition and $d=1$; hence the $1/\sqrt{\tau}$ scaling of the 
defect density is consistent with the Kibble-Zurek prediction. 

\begin{figure}[htb]
\begin{center} 
\includegraphics[height=2.2in,width=2.8in,angle=0]{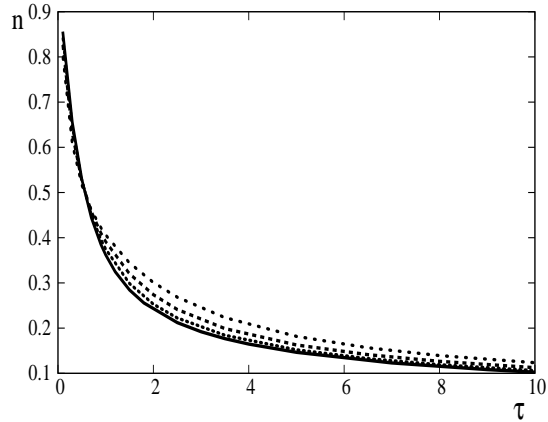}
\caption{Plot of kink density $n$ versus $\tau$ as obtained for $h =
0.2, 0.4, 0.6, 0.8$ (from bottom to top in the large $\tau$ region), with
$J_y= 1$ for anisotropic quenching. 
For large $\tau$, $n$ increases with increasing $h$, whereas for
small $\tau$, it decreases with increasing $h$.} \label{dmdsnvstau_jx}
\end{center} \end{figure}

On the other hand, if we look at the anisotropic quenching $J_x \sim 
t/\tau$, the off-diagonal terms of the Hamiltonian in (\ref{dmdsmatrix}) 
pick up a time dependence and hence a direct application of the LZ transition 
probability is not possible. To overcome this problem, we rewrite the 
Hamiltonian in the basis of the initial and final eigenstates when 
$J_x \to -\infty$ and $J_x \to \infty$. 
The eigenstates of the Hamiltonian in these limits are given by 
$$|e_{1k} \rangle = \sin (k/2) |0 \rangle + i\cos (k/2) |k,-k \rangle$$ and
$$|e_{2k} \rangle = \cos (k/2) |0 \rangle - i\sin (k/2) |k,-k \rangle,$$
with eigenvalues $\lambda_1 = t/\tau$ and $\lambda_2 = -t/\tau$ respectively;
the system is in the state $|e_{1k} \rangle$ initially. A general state vector
can be expressed as a linear combination of $|e_{1k} \rangle$ and $|e_{2k} 
\rangle$,
\begin{equation} |\psi_k(t) \rangle = C_{1k}(t) |e_{1k} \rangle + C_{2k}(t) 
|e_{2k} \rangle . \end{equation}
The initial condition in the anisotropic case is $C_{1k}(-\infty) = 1$ and
$C_{2k}(-\infty) = 0$. The unitary transformation to rewrite the Hamiltonian 
in the $|e_{1k} \rangle$ and $|e_{2k} \rangle$ basis is given by
$H'_k (t) = U^\dagger H_k (t) U$, where
\begin{eqnarray} U ~=~ \left[ \begin{array}{cc} \cos (k/2) & \sin (k/2) \\
- i \sin (k/2) & i \cos (k/2) \end{array} \right], \nonumber \end{eqnarray}
and the new Hamiltonian is
\begin{eqnarray} H'_k (t) &=& -~[ h ~+~ (J_x+J_y)\cos k] ~I_2 \nonumber \\
& & + \left[ \begin{array}{cc} 
J_x + J_y \cos 2k + h \cos k & J_y \sin 2k + h \sin k \\
J_y \sin 2k + h\sin k & -J_x - J_y\cos 2k - h \cos k \end{array} \right].
\label{dmdsmatrix2} \end{eqnarray}
By virtue of the unitary transformation, the time dependence is entirely 
shifted to the diagonal terms which makes it possible to apply the LZ 
transition formula. Evaluating the probability of a non-adiabatic 
transition for the mode $k$ and integrating over the Brillouin zone, it 
can be shown that the density of defects in the final state following the 
anisotropic quenching is given by \cite{dmdsmukherjee07}
\begin{eqnarray} n &\sim& \frac{4 J_y}{\pi \sqrt{\tau}(4J_y^2-h^2)}~~~~
{\rm for}~~~ h<2J_y, \nonumber \\
&\sim& \frac{h}{\pi \sqrt{\tau}(h^2-4J_y^2)}~~~~{\rm for}~~~ h>2J_y.
\label{dmdsaniso_defect} \end{eqnarray}
The scaling behavior of the density of defects with $\tau$ is shown in Fig. 
\ref{dmdsnvstau_jx}.

Although Eqs. (\ref{dmdsaniso_defect}) satisfy the Kibble-Zurek 
scaling, the density of defects diverges for $h=2J_y$, i.e., in a passage 
through the multicritical points. This necessitates a generalization of 
the Kibble-Zurek prediction for quenching through a multicritical point. 
In the next section, we propose a generalized scaling for the density of 
defects valid for quenching through a critical point as well as
a multicritical point. 

\begin{figure}[htb]
\begin{center} 
\includegraphics[height=2.2in,width=2.8in,angle=0]{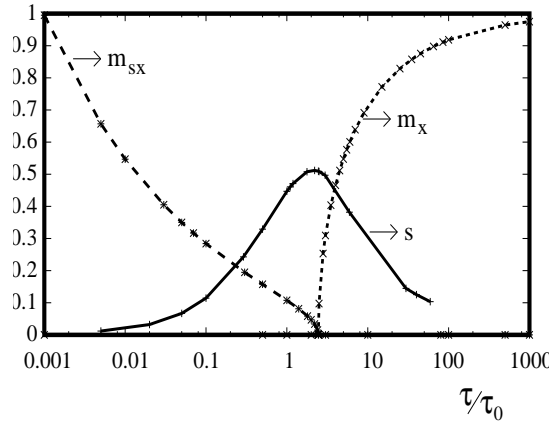}
\caption{Plot of von Neumann entropy density $s$, staggered magnetization
$m_{sx}$ and magnetization $m_x$ as a function of $\tau/\tau_0$, for
$J_y = 1$ and $h = 0.2$ in the anisotropic quenching case.} 
\label{dmdsentropy} 
\end{center} 
\end{figure}

In Fig. \ref{dmdsentropy}, we present the variation of the von Neumann 
entropy density of the final state defined by \cite{dmdslevitov06} 
\begin{eqnarray} s ~=~ -\int^\pi_0 \frac{dk}{\pi} ~[~p_k \ln (p_k) ~+~ 
(1 - p_k) \ln (1 - p_k) ~]. \end{eqnarray}
following the anisotropic quenching with the rate of quenching. Even 
though the final state is a pure state as a result of a unitary dynamics, 
it can also be viewed locally, or on a coarse-grained wave vector scale, as a 
decohered (mixed) state \cite{dmdslevitov06}. In the limit of $\tau \to 0$, 
the system does not get enough time to evolve; hence it largely retains its 
initial antiferromagnetic order which results in a low local entropy 
density. On the other hand, for slow quenching ($\tau \to \infty$), the 
system evolves adiabatically, always remaining close to its instantaneous 
ground state; this results in a final state with local ferromagnetic 
ordering, and hence again with a low local entropy density. We observe that 
the entropy density shows a maximum at a characteristic time scale $\tau_0$ 
where the magnetic ordering of the final state also changes from 
antiferromagnetic to ferromagnetic. 

Let us briefly mention a few other interesting variants of the transverse 
quenching scheme. In a repeated quenching dynamics of the same model,
the transverse field $h(=t/\tau)$ is quenched repeatedly between $-\infty$ 
and $+\infty$ \cite{dmdsmukherjee08}. We refer to a single passage from $h 
\to - \infty$ to $h \to +\infty$ or the other way around as a half-period of 
quenching. An even number of half-periods corresponds to the transverse
field being brought back to its initial value of $-\infty$, whereas, in the 
case of an odd number of half-periods, the dynamics is stopped at $h \to 
+\infty$. The probability of a non-adiabatic transition at the end of $l$ 
half-periods can be shown to follow the recursion relation
\begin{eqnarray} p_k(l) = (1 - e^{-2\pi\gamma}) - (1 - 2e^{-2\pi\gamma})
\left[ 1 - p_k(l - 1))\right] ,\end{eqnarray}
eventually yielding the simplified form of
\begin{eqnarray} p_k(l) &=& \frac{1}{2} - \frac{(1 - 2e^{-2\pi\gamma})^l}{2}.
\end{eqnarray}

For large $\tau$, the density of defects is generally found to vary as 
$1/\sqrt{\tau}$ for any number of half-periods. On the other hand, for 
small $\tau$, it shows an increase in kink density for even values of $l$.
However, the magnitude is found to depend on the number of half-periods of 
quenching. For two successive half-periods, the defect density is found 
to decrease in comparison to a single half-period, 
suggesting the existence of a corrective mechanism in the reverse path. 
The entropy density increases monotonously 
with the number of half-periods, and shows qualitatively the same behavior 
for any number of half-periods. For a large number of repetitions 
$(l \to \infty)$, the defect density saturates to the value $1/2$ while 
the local entropy density saturates to $\ln 2$.

\begin{figure}[htb]
\begin{center} \includegraphics[height=1.9in]{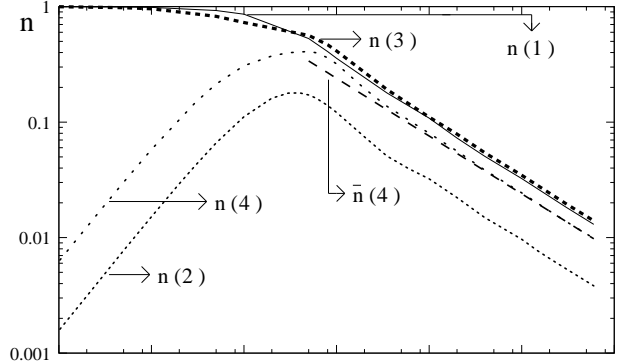}
\caption{Plot of kink density $n$ after $l$ half cycles as a function 
of $\tau$, for $J_x-J_y =1$ and $l = 1,2,3,4$. ${\bar n} (4)$ denotes the 
defect density as obtained from the analytical expression. In the limit of 
large $\tau$, the numerical results match perfectly with the analytical 
results.} \end{center} \end{figure}

The effects of interference on the quenching dynamics of Hamiltonian in 
(\ref{dmdsmatrix}) when the transverse field $h(t)$ varies sinusoidally with 
time as $h = h_0\cos{\omega t}$, with $|t| \leq \pi/\omega$ has also been 
studied in a recent work \cite{dmdsmukherjee09}. 
In this scheme of quenching, the time interval between two successive 
passages through the quantum critical points can be small enough for the 
presence of non-trivial effects of interference in the dynamics of the system.
It has been shown that for a single passage through a quantum critical point, 
the interference effects do not contribute leading to a kink density 
which goes as $n \sim \sqrt{\omega}$. On the other hand, repeated passages 
through the quantum critical points result in an oscillatory behavior of the 
kink density as well as the entropy density. 

\section{Quenching through a multicritical point}

We shall now propose a general scaling scheme valid for quenching through a 
multicritical point as well as a critical point \cite{dmdsdivakaran09} using
the LZ non-adiabatic transition probability \cite{dmdslandau,dmdssei} discussed 
before. We begin with a generic $d$-dimensional model Hamiltonian of the form
\begin{equation} H ~=~ \sum_{\vec k} ~\psi^{\dagger}(\vec k) ~[ \left( \lambda
(t) + b(\vec k) \right) ~\sigma^z ~+~ \Delta(\vec k) ~\sigma^+ ~+~ \Delta^* 
(\vec k) ~\sigma^- ]~ \psi(\vec k), \label{dmdsham} \end{equation}
where $\sigma^{\pm} = \sigma^x \pm i \sigma^y$, $b(\vec k)$ and $\Delta 
(\vec k)$ are model dependent functions, and $\psi(\vec k)$ denotes the 
fermionic operators $( \psi_1 (\vec k), \psi_2 (\vec k))$. The above 
Hamiltonian represents, for example, a one-dimensional transverse Ising or 
$XY$ spin chain \cite{dmdslieb61} as shown in Eq. (1.8), or an extended Kitaev 
model in $d=2$ written in terms of Jordan-Wigner fermions 
\cite{dmdskitaev,dmdssengupta08}. The excitation spectrum takes the form
\begin{equation} \epsilon_k^{\pm}=\pm \sqrt{(\lambda(t)+b(\vec k))^2+|
\Delta(\vec k)|^2}. \label{dmdsspectrum2} \end{equation}

We assume that the parameter $\lambda(t)$ varies linearly as $t/\tau$ and 
vanishes at $t=0$. The parameters $b(\vec k)$ and $\Delta (\vec k)$ are 
assumed to vanish at the quantum critical point in a power-law fashion given by
\begin{equation} b(\vec k)\sim |\vec k|^{z_1} ~~~{\rm and}~~~ \Delta(\vec k)
\sim |\vec k|^{z_2}, \label{dmdsexp} \end{equation}
where we have taken the critical mode to be at ${\vec k}_0=0$ without any 
loss of generality. Eq. (\ref{dmdsexp}) also implies 
that the system crosses the gapless point at $t=0$ when $\lambda=0$, and 
$b({\vec k})$ and $\Delta({\vec k})$ also vanish for the critical mode 
${\vec k} =0$. Many of the models described by Eq. (\ref{dmdsham}) exhibit 
a quantum phase transition with the exponents associated with the quantum 
critical point being $z_1>z_2$ and hence $z=z_2=1$. We shall however explore 
the more general case encountered at a multicritical point where the 
dynamically exponent $z$ is not necessarily given by $z_2$.

The Schr\"odinger equation describing the time evolution of the system when 
$\lambda$ is quenched is given by $i \partial \psi_{\vec k} /\partial t = 
H \psi_{\vec k}$, where we have 
once again defined $\psi_{\vec k}$ as $\psi_{\vec k} = \tilde C_{1,{\vec k}} 
|0 \rangle + \tilde C_{2,{\vec k}} |{\vec k},-{\vec k} 
\rangle$. Using the Hamiltonian in Eq. (\ref{dmdsham}), we can write
\begin{eqnarray} i \frac {\partial \tilde {C}_{1,\vec k}}{\partial t} &=& 
\left(\frac {t}{\tau} + b(\vec k)\right) ~ \tilde {C}_{1,\vec k} ~+~ 
\Delta(\vec k) ~ \tilde {C}_{2,\vec k}, \nonumber \\
i \frac {\partial \tilde {C}_{2,\vec k}}{\partial t} &=& -\left( \frac {t} 
{\tau} + b(\vec k)\right) ~\tilde {C}_{2,\vec k} ~+~ \Delta^*(\vec k)~ 
\tilde {C}_{1,\vec k}. \label{dmdsschrodinger} \end{eqnarray}
One can now remove $b(\vec k)$ from the above equations by redefining 
$t/\tau + b(\vec k) \to t/\tau$; thus the exponent $z_1$ defined in Eq. 
(\ref{dmdsexp}) does not play any role in the following calculations. 
Defining a new set of variables 
$C_{1,\vec k} = \tilde C_{1,\vec k} \exp (i \int^t dt' ~t'/\tau)$ and
$C_{2,\vec k} = \tilde C_{2,\vec k} \exp (-i \int^t dt' ~t'/\tau)$, 
we arrive at a time evolution equation for $C_1 (\vec k)$ given by
\begin{equation} \left(\frac {d^2}{dt^2} ~-~ 2i \frac {t}{\tau} 
\frac{d}{dt} ~+~ |\Delta (\vec k)|^2 \right)C_{1,\vec k} ~=~ 0. \end{equation}
Further rescaling $t \to t\tau^{1/2}$ leads to 
\begin{equation} \left(\frac{d^2}{dt^2} ~-~ 2i t \frac{d}{dt} ~+~ |
\Delta(\vec k)|^2 \tau \right)C_{1,\vec k} ~=~ 0. \end{equation}
If the system is prepared in its ground state at the beginning of the 
quenching, i.e., $C_1(\vec k)=1$ at $t=-\infty$, the above equation 
suggests that the probability of the non-adiabatic transition, $p_{\vec k} =
\lim_{t \to +\infty} |C_{1,\vec k}|^2$, must have a functional dependence on 
$|\Delta(\vec k)|^2\tau$ of the form
\begin{equation} p_{\vec k} ~=~ f(|\Delta(\vec k)|^2\tau). \end{equation}
The analytical form of the function $f$ is given by the general LZ 
formula \cite{dmdssei,dmdslandau}. The defect density in the final state is 
therefore given by \cite{dmdsdivakaran09}
\begin{equation} n ~=~ \int\frac{d^d k}{(2\pi)^d} ~f(|\Delta(\vec k)|^2\tau) ~
=~ \int \frac{d^d k} {(2\pi)^d} ~f(|\vec k|^{2z_2}\tau). \end{equation}
The scaling $|\vec k| \to |\vec k|^{2z_2}\tau$ finally leads to a scaling 
of the defect density given by 
\begin{equation} n ~\sim~ 1/\tau^{d/(2z_2)}. \label{dmdsdef} \end{equation}

Let us recall the quenching dynamics of the transverse $XY$ 
spin chain discussed before \cite{dmdslevitov06,dmdsmukherjee07}. 
When the system is quenched across the Ising or 
anisotropic critical line by linearly changing $h$ or $J_x$ as $t/\tau$, 
$\Delta (\vec k)$ vanishes at the critical point as $\Delta (\vec k) \sim |
\vec k|$ yielding $z_2=z=1$; hence the generalized scaling form given in Eq.
(\ref{dmdsdef}) matches with the Kibble-Zurek prediction with $\nu = z=1$ as
mentioned before. The situation, however, is different when the
system is swept across a multicritical point. 

\begin{figure}
\begin{center} \includegraphics[height=3.2in,angle=-90]{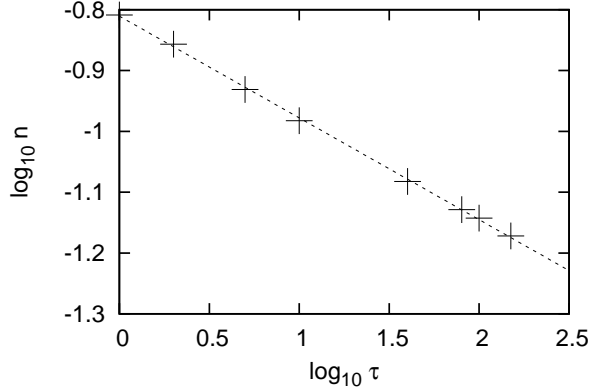}
\caption{$n$ vs $\tau$ obtained by numerically solving Eq. 
(\ref{dmdsschrodinger}) at the multicritical point of the $XY$ model in a 
transverse field, with $h=10$ and $J_y=5$. The line has a slope of $-0.16$.} 
\label{dmdsnvstau_multi} \end{center} \end{figure}

Putting $h=2J_y$ in the Hamiltonian in (\ref{dmdsmatrix2}),
let us analyze the scaling of the diagonal and off-diagonal terms of the 
Hamiltonian at the multicritical point. When $J_x=J_y$, the diagonal term of 
the Hamiltonian scales as $-J_y(\pi-k)^2$ whereas the off-diagonal term 
$|\Delta(k)|\sim|\pi-k|^3$ when expanded around the critical mode at $k=\pi$.
The dynamical exponent is obtained from the diagonal
term so that $z=z_1=2$. As discussed above, the off-diagonal term or more 
precisely the exponent $z_2$ determines the scaling of the defect density.
The exponent $z_2=3$ for the $XY$ multicritical point; hence the defect 
density scales as $1/\tau^{1/6}$. The density of defects obtained by numerical 
integration of the Schr\"odinger equation in (\ref{dmdsschrodinger}) as shown 
in Fig. \ref{dmdsnvstau_multi} supports this scaling behavior. For a 
non-linear quench across a multicritical point \cite{dmdsmondal091},
when the parameter $\lambda$ is quenched as $\lambda \sim (t/\tau)^{\alpha} 
sgn(t)$, both $z_1$ and $z_2$ come into play and the scaling of defect 
density gets altered to $n \sim \tau^{-d\alpha\nu/[\alpha(z_2\nu+1)+
z_1\nu(1-\alpha)]}$; this reduces to the form presented above for 
$\alpha=1$ and $z_2 \nu=1$.

\section{Quenching along a gapless line}

We now shift our focus to quenching the $XY$ Hamiltonian in (\ref{dmdsh2})
along the gapless line $h=J_x+J_y$ by varying the anisotropy parameter 
$\gamma=J_x-J_y$ as $t/\tau$, keeping $J_x+J_y$ fixed \cite{dmdsdivakaran08}.
For convenience, let us set $J_x+J_y=h=1$. The excitation gap vanishes along
this line for the mode $k=\pi$. We rewrite Eq. (\ref{dmdsmatrix}) with the 
present notation in the form
\begin{equation} H_k=\left[\begin{array}{ll} 
1+\cos k&~~~~ i \gamma \sin k \\
-i\gamma\sin k&~~~~-1-\cos k \end{array}\right]. \label{dmdsmatrix1} 
\end{equation}

\begin{figure}
\begin{center} \includegraphics[height=2.0in]{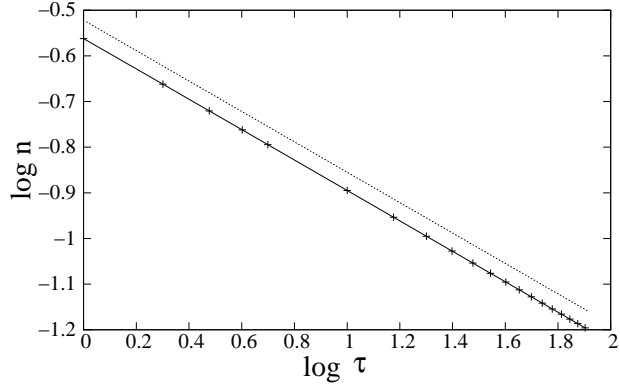}
\caption{$n$ vs $\tau$ obtained by numerically solving Schr\"odinger
equation with $\gamma(t) \sim t/\tau$ keeping $h=2J_y$. 
Also shown is a line with slope 1/3 for
comparison.} \label{dmdsnvstau_gapless} \end{center} \end{figure}

We once again encounter a situation in which the off-diagonal terms are 
time-dependent. Noting that the asymptotic form of the Hamiltonian at $t 
\to \pm \infty$ is given by
\begin{equation} H=\frac{t}{\tau} \sin k~ \hat\sigma^x, \end{equation}
we make a basis transformation to a representation in which $\sigma^x$ is 
diagonal. The Hamiltonian in (\ref{dmdsmatrix1}) then gets modified to 
the form \begin{equation} \left[\begin{array}{ll}
(t/\tau) ~\sin k& ~~~~1+\cos k\\
1+\cos k&~~~~-(t/\tau)~ \sin k \end{array}\right], \end{equation}
where the time dependence has been shifted to the diagonal 
terms only. Applying the LZ transition probability formula, the probability 
of excitations is found to be
\begin{equation} p_k=e^{-\pi \tau |\Delta_k|^2/\sin k}, \end{equation}
where $|\Delta_k|^2=|1+\cos k|^2=(\pi-k)^4/4$ when expanded about $k=\pi$.
Integrating $p_k$ over the Brillouin zone, we find the that density of 
defects falls off as $n\sim1/\tau^{1/3}$. Fig. \ref{dmdsnvstau_gapless}
which shows $n$ vs $\tau$ obtained by numerical integration
of the Schr\"odinger equation confirms the $\tau^{-1/3}$ behavior. 

It is to be noted that the Kibble-Zurek formalism cannot address defect 
generation along a gapless line. We will now present another general scaling 
argument for moving along a gapless line in a $d$-dimensional system. 

Let the excitations on the gapless quantum critical line be of the form 
$\omega_{\vec k} \sim \lambda | \vec k|^z$, where $z$ is the dynamical exponent
and the parameter $\lambda=t/\tau$ is quenched from $-\infty$ to $\infty$. 
Using a perturbative method involving the Fermi Golden rule along with the 
fact that the system is initially prepared in the ground state, the defect 
density can be approximated as \cite{dmdspolkovnikov07}
\begin{equation} n ~\simeq~ \int \frac {d^d k}{(2\pi)^d} ~\left|
\int_{-\infty}^\infty d\lambda ~\langle \vec k | \frac {\partial}{\partial 
\lambda} |0 \rangle ~e^{i\tau \int^{\lambda} \delta \omega_{\vec k} 
(\lambda') d\lambda'}\right|^2 . \end{equation}
Let us assume a general scaling form for the instantaneous excitation, $\delta
\omega_{\vec k} (\lambda') = k^a f(\frac{\lambda k^z}{k^a})$, where $k = |
\vec k|$ and $k^a$ denotes the higher order term in the excitation spectrum 
on the gapless line. Defining a new variable $\xi = \lambda k^{z-a}$, we 
obtain the scaling behavior of the defect density as \cite{dmdsdivakaran08}
\begin{equation} n ~\sim ~1/\tau^{d/(2a -z)} ~. \label{dmdsgeneral} 
\end{equation}
The case $d=1$, $a =2$ and $z=1$ has been discussed in this section.
Note that the correlation length exponent $\nu$ does not appear in the 
expression in Eq. (\ref{dmdsgeneral}) because our quench dynamics always
keeps the system on a critical line. The scaling for quenching along a 
gapless line is less universal in comparison to the general Kibble-Zurek 
prediction. For a non-linear quench \cite{dmdsmondal09} $\lambda \sim 
(t/\tau)^{\alpha} sgn(t)$, this scaling form gets modified to $n \sim 
\tau^{-d\alpha/[a(\alpha+1)-z]}$; this reduces to $1/\tau^{1/3}$
for $\alpha=d=z=1$ and $a=2$. 

In passing, let us introduce a variant of the $XY$ Hamiltonian with
an alternating field given by \cite{dmdsjhhp75,dmdsviola08}
\begin{eqnarray} H &=& - ~\frac{1}{2} ~[ ~\sum_j ~{(J_x + J_y)} (\sigma^x_j 
\sigma^x_{j+1} + \sigma^y_j \sigma^y_{j+1}) \nonumber \\ 
& & ~~~~~~~~+~ {(J_x-J_y)} (\sigma^x_j \sigma^x_{j+1} -\sigma^y_j 
\sigma^y_{j+1}) + (h-(-1)^j\delta ) \sigma^z_j]. \label{dmdsh1} \end{eqnarray}
The strength of the transverse field alternates between $h+\delta$
and $h-\delta$ on the odd and even sites respectively. For $\delta=0$, we 
recover the conventional $XY$ model in a transverse field. This Hamiltonian
can also be solved exactly by a Jordan-Wigner transformation by taking
care of the two underlying sublattices, i.e., by defining $a_k$ and
$b_k$ as two different Jordan-Wigner fermions. The Hamiltonian in the 
basis $(a_k^{\dagger}, a_{-k},b_k^{\dagger},b_{-k})$ can be written as
\begin{eqnarray} H_k=\left[ \begin{array}{cccc} 
h +J \cos k & i\gamma \sin k & 0 & -\delta \\
-i \gamma \sin k & - h - J \cos k & \delta & 0 \\
0 & \delta & J \cos k - h & i \gamma \sin k \\
-\delta & 0 & - i \gamma \sin k & - J \cos k + h \end{array} \right]. 
\label{dmdsdimerh2} \end{eqnarray}

The excitation spectrum of the above Hamiltonian is 
\begin{eqnarray} \Lambda_k^\pm &=& [~ h^2 +\delta^2 +J^2 \cos^2 k + 
\gamma^2 \sin^2 k \nonumber \\
& & \pm 2 \sqrt{h^2 \delta^2 + h^2 J^2 \cos^2 k + \delta^2 \gamma^2 \sin^2 
k}~ ]^{1/2}~, \label{dmdslam3} \end{eqnarray}
where $J=J_x+J_y$ and $\gamma=J_x-J_y$,
with four eigenvalues given by $\pm\Lambda_k^{\pm}$.
In the ground state, $-\Lambda_k^+$ and $-\Lambda_k^-$ are filled.
The vanishing of $\Lambda_k^-$ dictates the quantum critical point, and
the critical exponents are obtained by studying the behavior of
$\Lambda_k^-$ in the vicinity of the critical point.
The minimum energy gap in the excitation spectrum occurs at $k=0$
and $k=\pi/2$. The corresponding phase boundaries $h^2=\delta^2+J^2$ and
$\delta^2=h^2+\gamma^2$ signal quantum phase transitions from a paramagnetic 
to a ferromagnetic phase respectively.
We extend the study of quenching through a gapless phase in this Hamiltonian
by varying $\gamma$ as before along the phase boundary $h^2=\delta^2+J^2$.
On this gapless line, the dispersion of the low-energy
excitations at $k \to 0$ can be approximated as
\begin{equation} \Lambda_k^- ~=~ \sqrt{\frac{J^4 k^4}{4(\delta^2 + J^2)} + 
\frac{\gamma^2 J^2 k^2}{\delta^2 + J^2}} ~. \label{dmdslam6} \end{equation}
This suggests a truncation of the Hamiltonian in (\ref{dmdsdimerh2}) to a 
$2\times 2$ LZ Hamiltonian 
\begin{equation} h_k ~=~ \left[ \begin{array}{cc} 
{\tilde \gamma(t)} k & {\tilde J}^2 k^2 /2 \\
{\tilde J}^2 k^2 /2 & {-\tilde \gamma(t)}k \end{array} \right], 
\label{dmdsham2} \end{equation}
where $\tilde \gamma$ and $\tilde J$ are renormalized parameters given by 
$\tilde \gamma = \gamma J/\sqrt{\delta^2+J^2}$ and $\tilde J^2 = J^2/ 
\sqrt{\delta^2+J^2}$. The argument which justifies the truncation of a $4 
\times 4$ matrix to a $2 \times 2$ matrix is presented in Ref. 
\cite{dmdsdivakaran08}. The diagonal terms in Eq. (\ref{dmdsham2}) describe 
two time-dependent levels approaching each other linearly in time (since 
$\gamma = t/\tau$), while the minimum gap is given by the off-diagonal term 
${\tilde J}^2 k^2 /2$. The probability of excitations $p_k$ from the ground 
state to the excited state for the $k$-th mode is given by the LZ transition 
formula \cite{dmdslandau,dmdssei}
\begin{equation} p_k ~=~ \exp ~[-~\frac{2\pi \tilde J^4k^4}{8 k d\tilde 
\gamma(t)/dt}] ~=~ \exp~[-~\frac{\pi J^3k^3\tau}{4\sqrt{\delta^2+J^2}}], 
\label{dmdspk} \end{equation}
and we get back the $\tau^{-1/3}$ scaling form of the defect density
as discussed above.

\section{Conclusions}

The non-equilibrium dynamics of a quantum system driven through a quantum 
critical point is indeed an exciting and fascinating area of research. 
The possibility of experimental realizations of quenching dynamics adds to 
the importance of the theoretical research. In this review we have discussed 
the scaling relations for the defect density in the final state following
a slow quench across a multicritical point and along a gapless line. In both 
cases, the relations obtained here using the LZ transition formula are not 
directly derivable using the Kibble-Zurek argument. But in all the cases, the
defect density scales as a power law with the rate of quenching. It should be 
noted that the idea of a dominant critical point has also been invoked to 
justify the scaling along a gapless line in one-dimensional spin models 
\cite{dmdsviola08}. A generalization of the studies presented here to 
higher dimensional quantum systems and to systems with quenched disorder are
some of the future directions of this subject. 

\vskip .8 cm
\begin{center}
\bf Acknowledgements
\end{center}

The authors acknowledge the organizers of the conference ``Quantum Annealing
and Quantum Computation". We also thank A. Polkovnikov, S. Sachdev,
G. E. Santoro and K. Sengupta for useful discussions at different occasions. 
AD also acknowledges R. Moessner and hospitality of MPIPKS, Dresden, where some part
of the work discussed in the review was done.

\end{document}